\documentclass[12pt]{JHEP}
\usepackage{graphics}
\usepackage{cite}
\newcommand{\beq}{\begin{equation}}
\newcommand{\eeq}{\end{equation}}
\newcommand{\bea}{\begin{eqnarray}}
\newcommand{\eea}{\end{eqnarray}}
\newcommand{\Z}{\cal Z}
\renewcommand{\d}{\delta}

\renewcommand{\b}{\beta}
\renewcommand{\a}{\alpha}

\newcommand{\n}{\nu}
\newcommand{\m}{\mu}

\newcommand{\s}{\sigma}
\newcommand{\D}{\Delta}

\newcommand{\tS}{\tilde{S}}
\renewcommand{\P}{{\cal P}}
\renewcommand{\th}{\theta}

\newcommand{\tU}{\tilde{U}}
\newcommand{\e}{\epsilon}

\newcommand{\oh}{\frac{1}{2}}

\newcommand{\dg}{\dagger}
\newcommand{\non}{\nonumber}

\newcommand{\rf}[1]{(\ref{#1})}

\title{What are the Confining Field Configurations of Strong-Coupling
Lattice Gauge Theory?}

\author{Manfried Faber \\ Inst. f\"ur Kernphysik, 
Technische Universit\"at Wien, A-1040 Vienna, Austria \\
E-mail: \email{faber@kph.tuwien.ac.at}}

\author{Jeff Greensite \\ Physics and Astronomy Department, San Francisco
State University, San Francisco, CA 94117 USA \\
E-mail: \email{greensit@quark.sfsu.edu}\\ Theory Group, Lawrence 
Berkeley National Laboratory, Berkeley, CA 94720 USA \\
E-mail: \email{JPGreensite@lbl.gov} }

\author{{\v S}tefan Olejn\'{\i}k \\ Institute of Physics, Slovak Academy 
of Sciences, SK-842 28 Bratislava, Slovakia \\
Email: \email{fyziolej@savba.sk}}

\abstract{Starting from the strong-coupling SU(2) Wilson action in
$D=3$ dimensions, we derive an effective, semi-local action on a
lattice of spacing $L$ times the spacing of the original lattice.  It
is shown that beyond the adjoint color-screening distance, i.e.\ for
$L \ge 5$, thin center vortices are stable saddlepoints of the
corresponding effective action.  Since the entropy of these stable
objects exceeds their energy, center vortices percolate throughout
the lattice, and confine color charge in half-integer representations
of the SU(2) gauge group.  This result contradicts the folklore that
confinement in strong-coupling lattice gauge theory, for $D>2$
dimensions, is simply due to plaquette disorder, as is the case in
$D=2$ dimensions.  It also demonstrates explicitly how the emergence
and stability of center vortices is related to the existence of color
screening by gluon fields.}


\preprint

\begin{document}


   Quark confinement is commonly attributed to the influence of some
special class of gauge field configurations, which dominate the QCD
vacuum at large scales.  Because of their high probability, these
dominant configurations would most naturally correspond to the
saddlepoints of an infrared effective action, derived at large scales
by integrating over high-frequency modes.  In strong-coupling lattice
gauge theory there are methods available which enable us to compute
the QCD spectrum and string tension analytically, and the same methods
could also be applied to extract a long-range effective action.  An
interesting question, then, is what type of saddlepoint configurations
are actually found at strong lattice couplings; it is likely that the
answer would also shed some light on QCD in the continuum limit.
 
   The classical Euclidean action of pure SU(N) gauge 
theory is stationary, or nearly so, at multi-instanton configurations.
In quantized lattice gauge theory,
however, we can imagine performing renormalization-group (RG)
transformations so as to obtain an effective action at some scale $R$.
For scales $R$ well below the confinement scale, the main effect of
the RG transformations will simply be the running of the lattice
coupling constant.  At larger scales, however, so-called irrelevant
operators can become important in the effective action.  As a
consequence, at these larger scales, the effective theory may have
non-trivial saddlepoints which are something other than instantons.

   There are good reasons to believe that at sufficiently large
scales, these non-trivial saddlepoints are center vortices.  On the
theoretical side, we note that the asymptotic string tension between
static color charges in SU(N) gauge theory depends only on the
$N$-ality of the color charge representation.  Although this fact is
deduced rather trivially from the possibility of color-screening by
gluon fields, it has some profound implications for the infrared
structure of the QCD vacuum.  Consider, for example, Wilson loops
$W_j(C)$ in SU(2) gauge theory, where $j=0,\oh,1,{3\over 2},...$
labels the group representation.  Wilson loop expectation values can
be viewed as a probe of vacuum fluctuations in the \emph{absence} of
external sources (think of evaluating a spacelike loop in the
Hamiltonian formulation), and large Wilson loops are presumed to
become ``disordered,'' i.e.\ have an area-law falloff, due to
averaging over certain types of large-scale fluctuations which
dominate the vacuum state.  Whatever the nature of these confining
fluctuations, they must have the highly non-trivial property of
disordering only the $j=$ half-integer loops, but not the $j=$ integer
loops.  Center vortices are the only configurations we know of that
have this property.  Dual-superconductor models, in which all
multiples of abelian electric charge (identified in an
abelian-projection gauge) are confined by the dual Meissner effect, do
not seem satisfactory.  In these models, the potential between charged
objects is roughly proportional to the electric charge.  But this
charge dependence cannot be correct, since gluons carrying two units
of electric charge are available to screen multiply charged sources,
and numerical simulations indicate that only odd multiples of the
abelian charge (non-zero $N$-ality) are confined, while even multiples
of abelian charge (zero $N$-ality) are screened \cite{j3}.  Center
vortices seem to be the natural way of accounting, in terms of
dominant field configurations, for this very fundamental distinction
between zero and non-zero $N$-ality, which is evident even in the
abelian projection.

   On the numerical side, there is now abundant evidence in favor of
the vortex theory of confinement
\cite{indirect,Zako,Jan98,mog,dFE,dFE1,Lang1,Tubby, bertle,ITEP,
ET,Hart}, which we will not attempt to review here.  The present
situation can just be summarized as follows: There exists a method
(known as ``center projection'') for locating center vortices on
thermalized lattices; the rationale undelying this method is explained
in ref.\ \cite{vf}.  By locating the vortices, their effects on
gauge-invariant observables such as Wilson loops, Polyakov lines,
topological charge, etc., can be studied in detail.  The numerical
evidence indicates that fluctuations in vortex linking number are the
origin of the asymptotic string tension of Wilson loops.  The free
energy of a vortex, inserted into a finite lattice via twisted
boundary conditions, has also been computed, and has been shown to
fall off exponentially with the lattice size at just the rate
predicted by the vortex theory \cite{ET,Hart}.

   Vortices presumably have a finite thickness comparable to the
adjoint string-breaking length, at about $1.25$ fm \cite{dFP}, where
the crossover from Casimir scaling to $N$-ality confinement occurs
(cf.\ the discussion in ref.\ \cite{Cas}).  Independent estimates of
the vortex thickness are based on measurements of ``vortex-limited''
Wilson loops in ref.\ \cite{Jan98}, and on the vortex free energy in
finite volumes \cite{ET}.  Both of these estimates put the vortex
thickness at a little over one fermi. Beyond this scale, the presence
of vortex sheets in the vacuum should be very evident.  A reasonable
conjecture is that if the appropriate effective action could be
determined at this scale, it would be found to have stable
saddlepoints corresponding to vortex configurations, which percolate
through the lattice according to the usual energy-entropy arguments.
Unfortunately, the calculation of long-range effective actions is very
difficult even numerically, via Monte Carlo RG methods, and at such
large scales the problem is quite intractable by perturbative (e.g.\
one-loop) methods.\footnote{On the other hand, there do exist some
intriguing results at one-loop that should be noted.  Diakonov
\cite{Mitya} has computed a one-loop effective potential for magnetic
flux tubes, and his result indicates that the potential is minimized
for magnetic flux in the center of the gauge group.  There is also the
old, but still provocative, Copenhagen vacuum picture, which is again
based on one-loop considerations \cite{CopVac}.}

   We therefore turn our attention, in this article, to
strong-coupling lattice gauge theory, where analytic methods can be
brought to bear at arbitrarily large scales.  In the strongly coupled
theory in $D>2$ dimensions, we have a linear static potential for all
color charge representations up to a screening scale of about four
lattice spacings (for the SU(2) Wilson action).  Beyond that scale,
the string tension depends only on the $N$-ality of the
representation, just as in the continuum theory.  Thus, if our
conjecture is correct, and if the $N$-ality dependence implies a vortex
mechanism, the effective strong-coupling action at a scale
beyond four lattice spacings should have saddlepoints which are stable
center vortices.

   There is, however, some folklore to the effect that confinement in strong
coupling lattice gauge theory, in $D>2$ dimensions, is essentially
the same as in $D=2$ dimensions, where the mechanism is simply
plaquette disorder.  If that were so, then vortices (or any other
topological objects) are irrelevant at strong couplings in any dimension.
This folklore is quite misleading, according to an argument presented in 
ref.\ \cite{j2}, which we now review.

\FIGURE[h!]{
\centerline{\scalebox{0.37}{\rotatebox{270}{\includegraphics{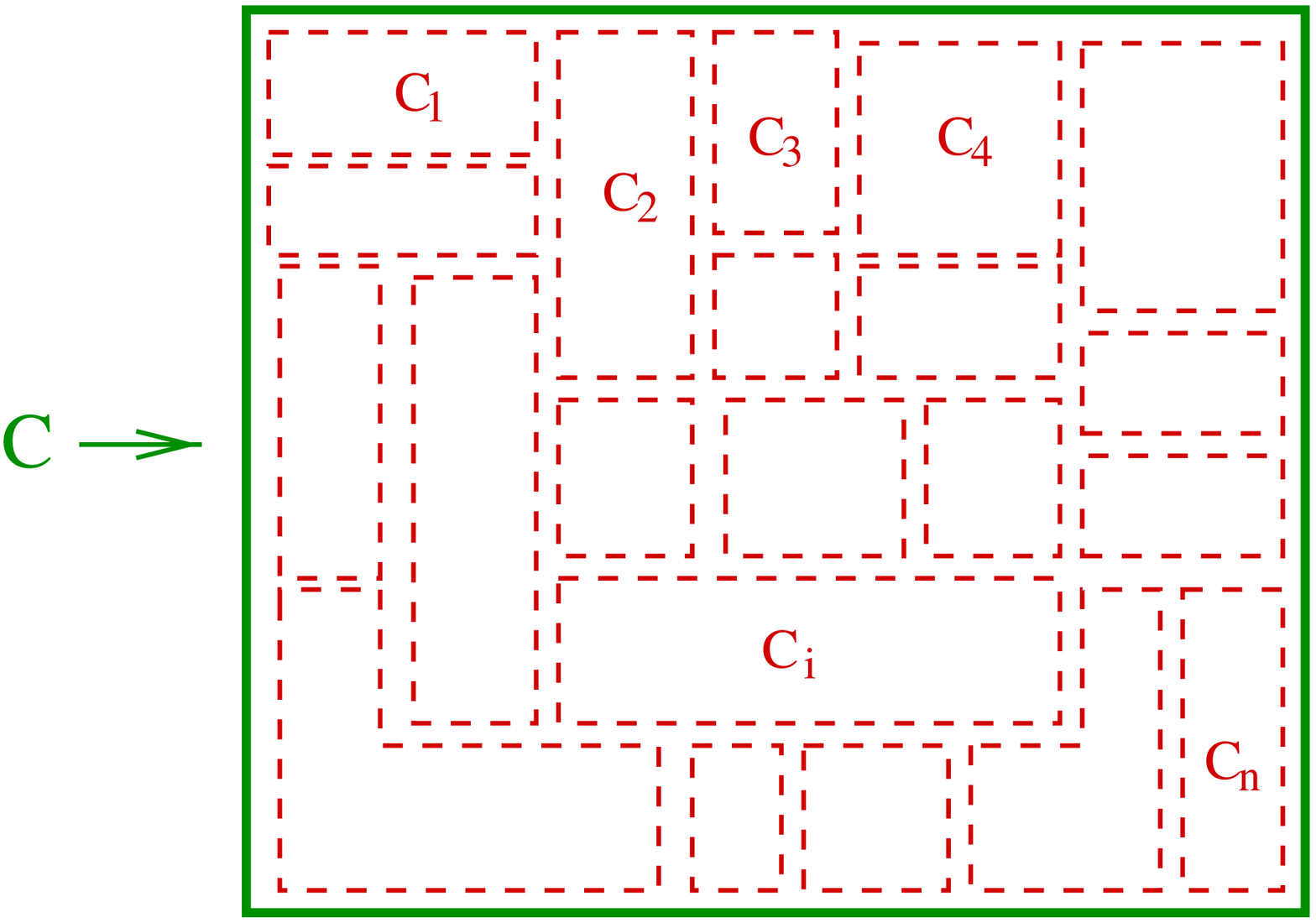}}}}
\caption{Decomposition of a large area bounded by planar loop $C$ into
many sub-areas bounded by planar loops $\{C_i\}$}
\label{loops}
}

   Consider SU(2) lattice gauge theory at strong-coupling, and denote
by $U(C)$ the product of link variables around loop $C$.  Let the
minimal area of a planar loop be decomposed into a set of smaller
areas, bounded by loops $\{C_i\}$, as shown in Fig.\ \ref{loops}.  
The area-law of a Wilson loop is thought to be due to ``magnetic disorder'', 
in which the gauge field strength fluctuates independently in sub-areas
of the minimal surface of the loop.  If this is so, then the holonomies
$\{U(C_i)\}$ should be (nearly) uncorrelated, for large
areas and small $\beta$.  The test for such independent fluctuation 
in the subareas $A(C_i)$ is whether
\beq
      <\prod_i F[U(C_i)]> \stackrel{?}{=}  \prod_i <F[U(C_i)]> 
\label{test}
\eeq
for any class function
\beq
        F[g] = \sum_{j\ne 0} f_j \chi_j[g]
\eeq
In fact, in $D=2$ dimensions, it is easy to show that this equality
is satisfied exactly.  However, for dimensions $D>2$, evaluating the
left- and right-hand sides of \rf{test} one finds for the exponential
falloff on each side \cite{j2}
\beq
  e^{-4\s {\cal P}(C)} \prod_i {1\over 3} f_1 \gg \prod_i f_1 e^{-4\s 
         {\cal P}(C_i)}
\eeq
where the inequality holds for perimeters ${\cal P}(C)\ll \sum_i 
{\cal P}(C_i)$.
The conclusion is that the holonomies $U(C_i)$ do \emph{not} fluctuate 
independently, even at strong-coupling, for $D>2$.

   If the loop holonomies in sub-areas of the loop are correlated,
then where is the magnetic disorder required to give an area-law
falloff for Wilson loops?  The question is resolved by extracting a
center element from the holonomies
\beq
        z[U(C_i)] = \mbox{signTr}[U(C_i)] \in Z_2
\eeq
and asking if these center elements fluctuate independently; i.e
\beq
      <\prod_i z[U(C_i)]> \stackrel{?}{=}  \prod_i <z[U(C_i)]> 
\eeq
In fact, it is easy to show that they do:
\beq
  e^{-\s A(C)} \prod_i {3\over 4\pi} = \prod_i {3\over 4\pi} e^{-\s A(C_i)}
\eeq
Thus, \emph{magnetic disorder is center disorder} in $D>2$ dimensions, 
at least at strong couplings.  Confining configurations must disorder 
the center elements $z$, but not the coset elements, of SU(2) holonomies
$U(C_i)$.  Again, the only configurations known to have this 
property are center vortices.  

   We now return to our conjecture that vortices are stable saddlepoints
of a long-range effective action.  Actually, there are various ways 
of integrating out the smaller-scale
fluctuations, to obtain an effective action at a larger scale.  One
simple approach is to superimpose, on a lattice of spacing $a$ with 
link variables denoted $U$, a lattice of spacing $La$ with links 
denoted $V$.  An effective action for the lattice with the larger
spacing can then be obtained from
\beq
  \exp\Bigl[S_{eff}[V]\Bigr] = \int DU \prod_{l'} 
\d\Bigl[V^\dg_{l'}(UU..U)_{l'}-I\Bigr] e^{S_W[U]}
\label{S1}
\eeq
where $(UU..U)_{l'}$ is the product of $U$-link variables along the
link $l'$ of the V-lattice, and $S_W$ is the Wilson action.  Obviously,
all observables computed on the V-lattice with $S_{eff}$ will agree with the 
corresponding quantity computed on the U-lattice using $S_W$.  

  It is trivial to compute $S_{eff}$ in $D=2$ dimensions, and the
result is
\bea
   \lefteqn{\exp\Bigl[S_{eff}[V]\Bigr]} 
\non \\ &\propto& \prod_{P'} \sum_j 
           (2j+1) \Bigl( I_{2j+1}(\b)\Bigr)^{L^2} \chi_j[V(P')]
\non \\
       &=& \exp\left[ \sum_{P'}\log\left( 1 + \sum_{j=\oh,1,{3\over 2}}
  (2j+1)\left({I_{2j+1}(\b)\over I_1(\b)}\right)^{L^2} \chi_j[V(P')] \right) 
        ~+~ \mbox{const.} \right]
\non \\
 &\approx&   \exp\Bigl[ 2\left({\b\over 4}\right)^{L^2}
                \sum_{P'} \chi_{1/2}[V(P')]   ~+~ \mbox{const.} \Bigr]
\label{S2}
\eea
where $V(P')$ is the product of V-links around the plaquette $P'$.
One might imagine that the action \rf{S2} is also a good approximation
in $D>2$ dimensions, at least at strong couplings, since this action
gives the correct string tension for fundamental representation Wilson
loops.  But a quick calculation of higher representation loops shows
that \rf{S2} cannot even be approximately correct for large $L$.  A loop
in the adjoint representation, for example, calculated on the U-lattice
with the Wilson action, is easily seen to have
an asymptotic perimeter law falloff
\beq
     W_{adj}(C) \sim \exp[-\m \P(C)]
\eeq
where
\beq
      \m = - 4 \log\left( {\b\over 4} \right)
\eeq
is the ``gluelump'' mass (gluon bound to a static adjoint color charge),
and $\P(C)$ is the loop perimeter in units of the U-lattice spacing.  However,
carrying out the same calculation with the effective action \rf{S2},
one finds instead (with $\P(C)$ again in U-lattice units) 
the erroneous result
\beq
      \m = - 4 L \log\left( {\b\over 4} \right) ~~~~~~~~ \mbox{(wrong)}
\label{wrong}
\eeq   
with an $L$-dependent gluelump mass.
The mismatch is not resolved by including a few more contours in the
effective action, so long as the coupling associated with each contour
is of order $(\b/4)^A$, where $A$ is the minimal area of the contour
in U-lattice units.

\FIGURE[h!]{
\centerline{\scalebox{0.50}{\includegraphics{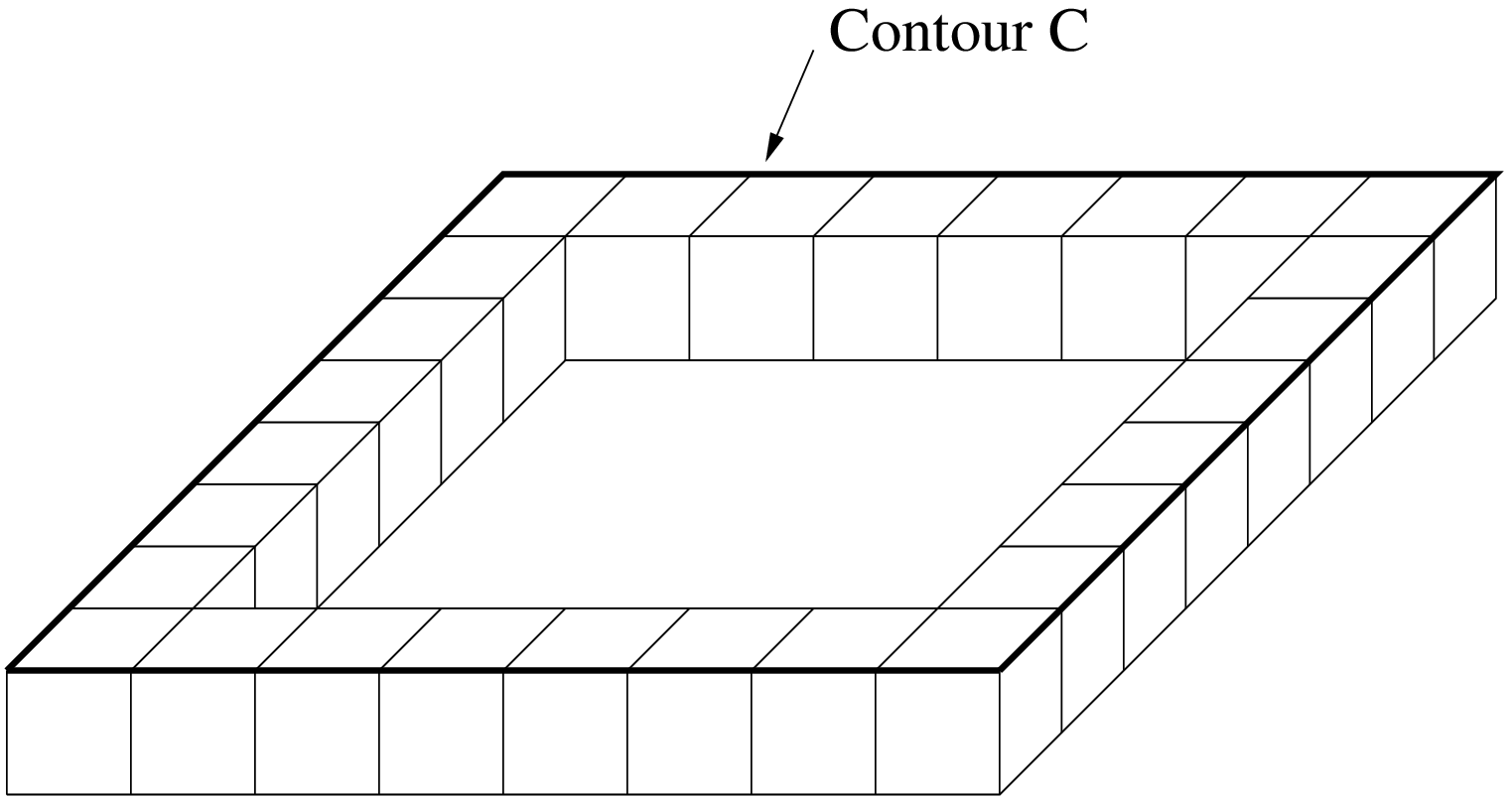}}}
\caption{Arrangement of U-plaquettes in a tube, around a rectangular
contour $C$ on the V-lattice.}
\label{fig0}
}

   In fact, what happens in $D>2$ dimensions is that the effective
action contains Wilson loops of all sizes in $j=$ integer representations,
and these loops are only suppressed by perimeter-law coefficients.
This is easily seen by bringing down a ``tube'' of plaquettes 
from
$\exp(S_W)$ in eq.\ \rf{S1}, such that the tube
borders a rectangular contour $C$ on the V-lattice, as shown in 
Fig.\ \ref{fig0}.  Integrating over all U-links 
in the tube, except those which lie on contour $C$, will yield 
contributions to $S_{eff}$ such as
\bea
    \exp\Bigl[S_{eff}[V]\Bigr] &\supset& \int DU_{l\in C}
\prod_{l'\in C} \d\Bigl[V^\dg_{l'}(UU..U)_{l'}-I\Bigr]    
\left({\b \over 4}\right)^{4(\P(C)-4)} \Bigl(\chi_{\oh}[U(C)]\Bigr)^2
\non \\
         &\supset& \left( {\b\over 4}\right)^{4(\P(C)-4)} 
    (\chi_1[V(C)] ~+~ \mbox{const.} )
\eea
This shows that $S_{eff}[V]$ contains adjoint (and, in general,
integer) representation loops with perimeter-falloff coefficients.
Such non-local terms introduce non-local correlations among
$SU(2)/Z_2$ coset elements in loop holonomies $\{U(C_i)\}$.
Truncating $S_{eff}[V]$ by removing these large loops will yield
erroneous results for any Wilson loop in representation $j>\oh$.

\FIGURE[h!]{
\centerline{\scalebox{0.50}{\includegraphics{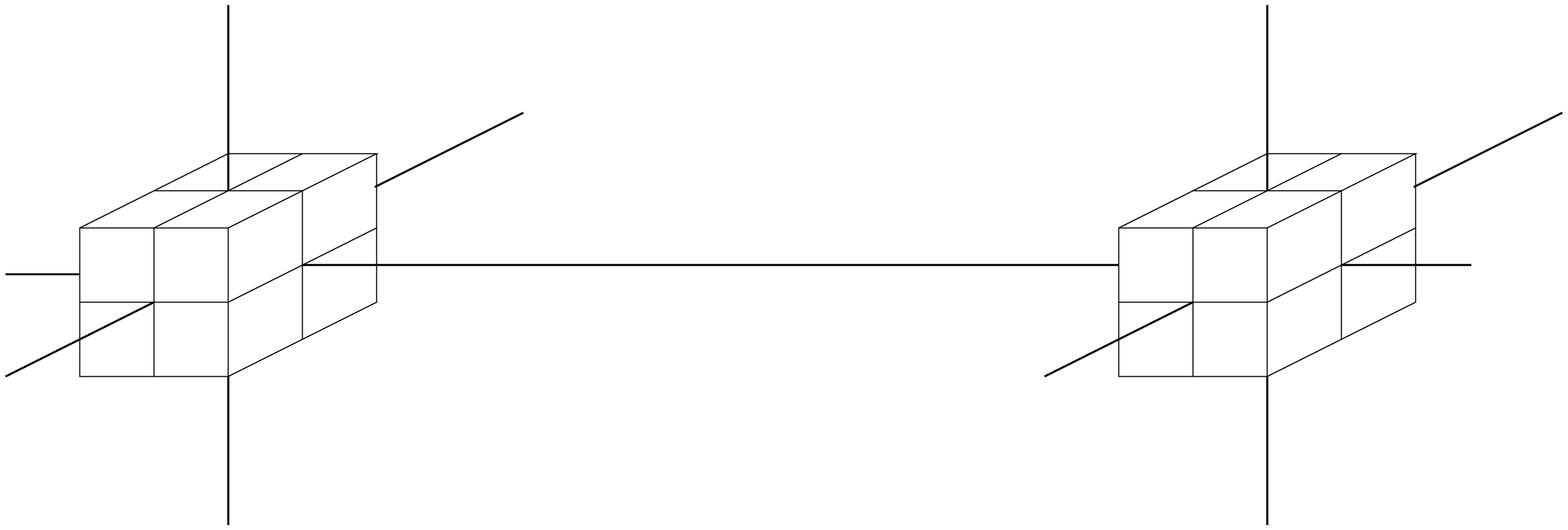}}}
\caption{The degrees of freedom in $\tS_L[V,\tU]$.}
\label{fig1}
}

   Our aim is to modify the prescription \rf{S1} for the effective
action, in such a way that at least the \emph{leading} contribution to
any Wilson loop on the V-lattice is obtained from a local action.  The
strategy for doing this is to prevent the formation of closed tube
diagrams, of the form shown in Fig.\ \ref{fig0}, bordering contours on
the V-lattice.  This can be accomplished by not integrating, in eq.\
\rf{S1}, over the U-links in a cube of volume $2^D$ around each site
on the V-lattice.\footnote{Non-local terms in the effective action
will still arise from closed tubes which go around the 2-cubes.
These, however, are associated with sub-leading contributions to color
screening; the leading contributions in $\b$, arising from diagrams
like Fig.\ \ref{fig0}, will now be generated by local terms.}  U-links
belonging to these 2-cubes will be denoted $\tU_l$. In order to ease
the task of illustration, we will work in $D=3$ dimensions, although
the extension to higher dimensions should be straightforward.  We then
have
\bea
      Z &=& \int DV \int \prod_{l\in 2-cubes} d\tU_l ~ 
                      \exp\Bigl[\tS_L[V,\tU]\Bigr]
\non \\
        &=& \int DV \int \prod_{l\in 2-cubes} d\tU_l 
     \left\{ \int \prod_{l''\not\in 2-cubes} dU_{l''} 
 \prod_{l'} \d\Bigl[V^\dg_{l'}(UU..U)_{l'}-I\Bigr] e^{S_W[U]} \right\}
\eea
where the long-range action $\tS_L[V,\tU]$ depends on the 
$V$-link variables, and on the $\tU$-links in 2-cubes around sites of the
V-lattice, as shown in Fig.\ \ref{fig1}.
 
   Now introduce, in each 2-cube, a set of plaquette variables
$\{h_{ij},g_{ij}\}$ which are Wilson lines beginning and ending at the
center of the 2-cube, and running around one of the plaquettes in the
cube.  The $h_{ij}$ lines run around plaquettes on the faces of the
2-cube, and the $g_{ij}$ lines run around plaquettes on the interior of
the 2-cubes.  To fix notation: The orientation of the $g_{ij}$ lines
is taken to be counterclockwise in either the $xy$-plane ($i=1$), the
$yz$-plane ($i=2$), or the $zx$-plane ($i=3$).  The second index ($j=1-4$)
distinguishes between the four interior plaquettes in a given plane
with the convention shown in Fig.\ \ref{fig5}, where
$(x_a,x_b)=(xy),(yz),(zx)$.  Each $h_{ij}$ line begins at the center
of the 2-cube, runs to a center of one of the faces of the 2-cube,
goes around one of the plaquettes on the face, and returns to the
center of the 2-cube.  The orientation around the $h$-plaquettes is
defined by a right-hand rule: the thumb points in an outward direction
normal to the 2-cube.  The first index $i=1,2,3$ refers to a face of
the 2-cube in the $xy$, $yz$, $zx$-planes, one lattice spacing away from the
center of the 2-cube, in the negative $z,x,y$ directions, respectively.
Index values $i=4,5,6$ refer to faces in the $xy$, $yz$, $zx$-planes one
lattice spacing away from the center in the positive $z,x,y$ directions,
respectively.  These conventions are illustrated in Fig.\ \ref{fig4}.

\FIGURE[h!]{
\centerline{\scalebox{0.30}{\includegraphics{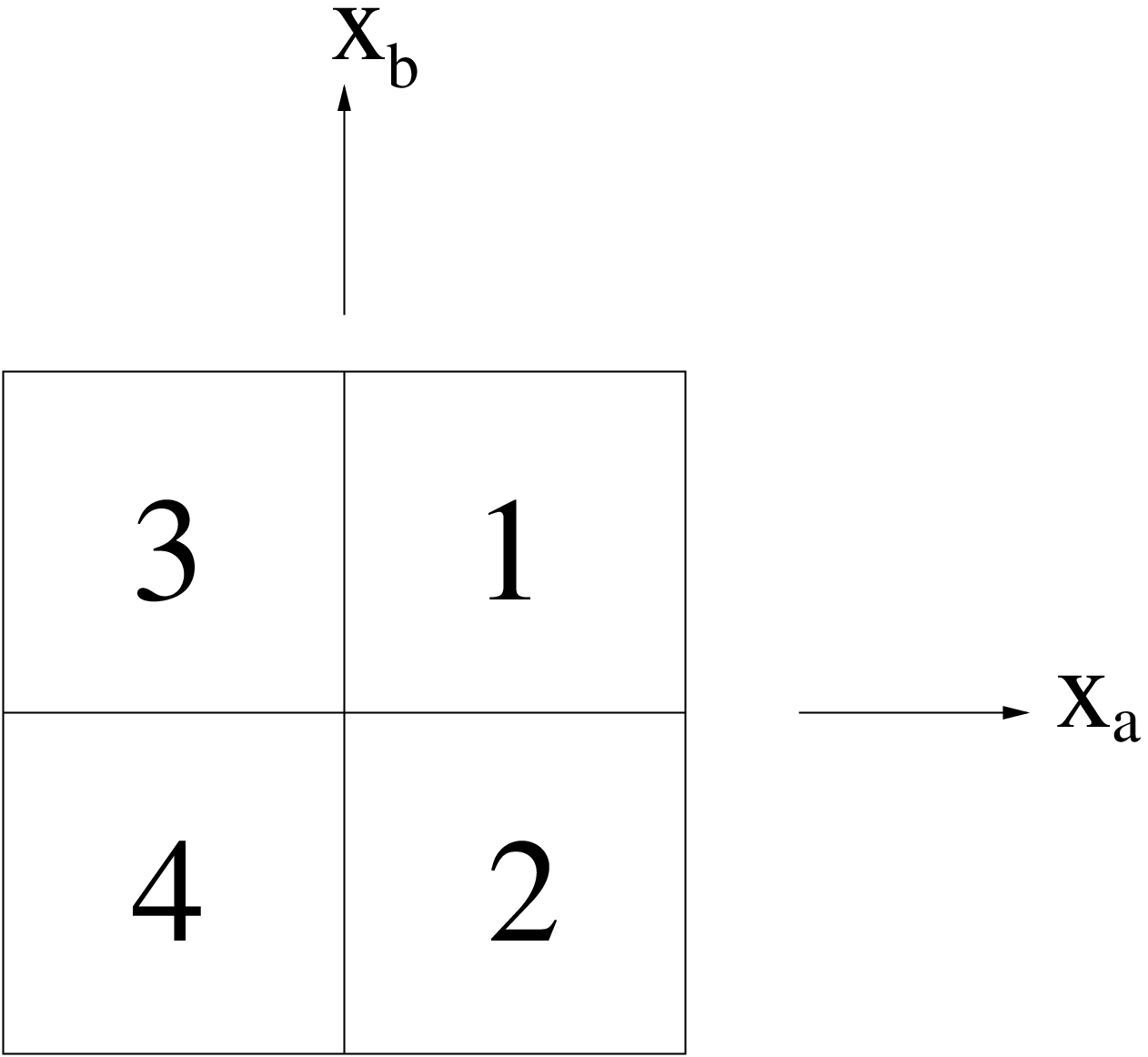}}}
\caption{Plaquette numbering convention.}
\label{fig5}
}

\FIGURE[h!]{
\centerline{\scalebox{0.50}{\includegraphics{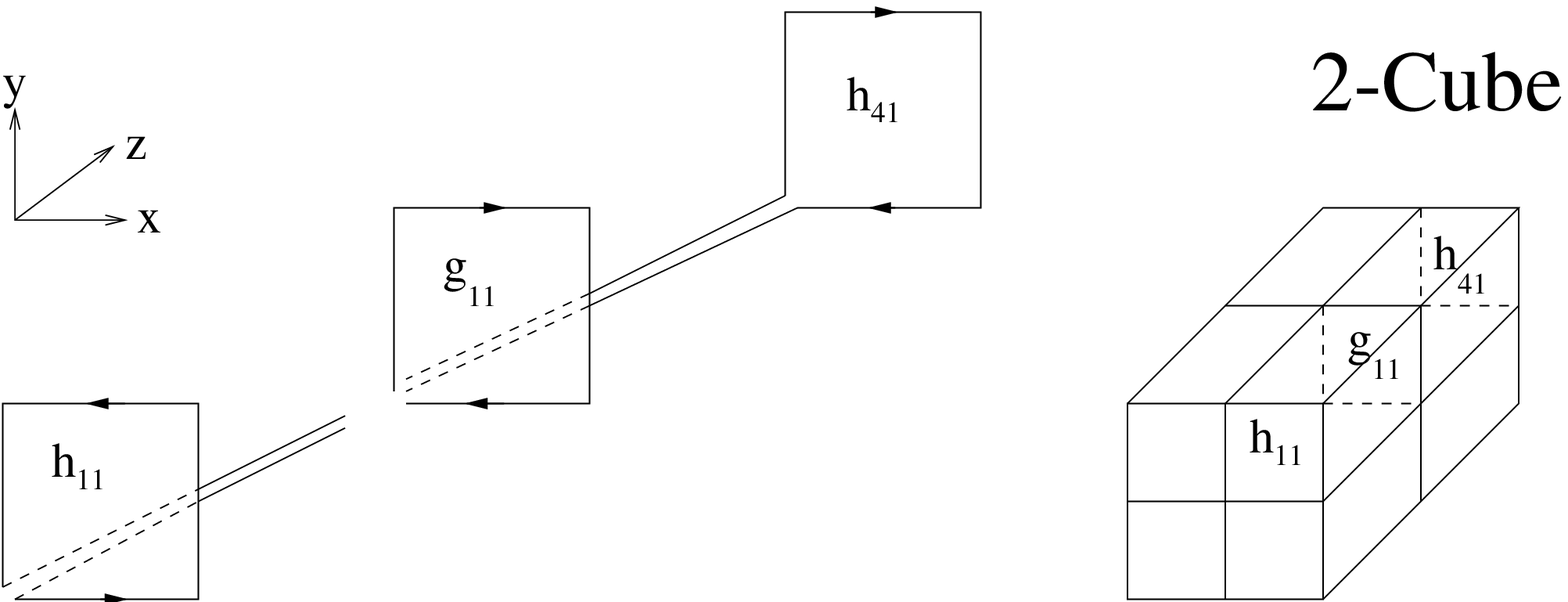}}}
\caption{Three plaquette variables on the 2-cube.}
\label{fig4}
}

   We then integrate over the $U$ links which do not belong to the
2-cubes.  Keeping, for each type of contribution, only terms of leading 
order in $\b$, the result is approximately
\bea
   Z &\approx& \int DV D\tU ~ \exp\left[ {\b\over 2} \sum(\mbox{Tr}[h] 
                         + \mbox{Tr}[g]) \right. 
\non \\
    & & + 2\left(\b\over 4\right)^{4(L-2)}\sum_{l'}
  f_{l'}^{ijkl} \mbox{Tr}[h^\dg_{ij}V_{l'}
            h^\dg_{kl}V^\dg_{l'}]
\non \\
   & & \left. + 2\left({\b\over 4}\right)^{L^2-4} \sum_{P'}
              \mbox{Tr}[VgVgV^\dg g^\dg V^\dg g^\dg] \right]
\eea
Coefficients $f_{l'}^{ijkl}=1$ if plaquette variables $h_{ij}$
and $h_{kl}$ on nearest-neighbor 2-cubes can be joined by a cylinder
of plaquettes, bordering link $l'$, of length $L-2$ U-lattice spacings.
Otherwise, $f_{l'}^{ijkl}=0$.

  The next step is to change integration variables from links $\tU$ to
plaquettes $g,h$.  This change of variables on the lattice was worked
out many years ago by Batrouni \cite{Batrouni}, and the result is simply
to introduce a Bianchi constraint into the integration measure\footnote{There
are 36 $g,h$ plaquette variables on the 2-cube, and 8 Bianchi constraints,
leaving 28 independent group-valued variables.  Similarly, up to 26 out of 54
link variables on the 2-cube can be gauge fixed to the identity, again leaving
28 independent group-valued variables.}
\bea
   Z &\approx& \int DV Dh Dg \prod_{2-cubes~K}~ \prod_{c\in K}
                   \d[\mbox{Bianchi}(c(K))]
\non \\
  & & \exp\left[ {\b\over 2} \sum(\mbox{Tr}[h]  + \mbox{Tr}[g]) \right.  
\non \\
    & & + 2\left(\b\over 4\right)^{4(L-2)}\sum_{l'}
  f_{l'}^{ijkl} \mbox{Tr}[h^\dg_{ij}V_{l'}
            h^\dg_{kl}V^\dg_{l'}]
\non \\
  & & \left. + 2\left({\b\over 4}\right)^{L^2-4} \sum_{P'}
              \mbox{Tr}[VgVgV^\dg g^\dg V^\dg g^\dg] \right]
\label{Z1}
\eea
Each 2-cube $K$ contains eight unit sub-cubes; the index $c$ in eq.\
\rf{Z1} labels these subcubes. The $\d$-function constraints force a
certain product of three $g$ and three $h$ variables on each subcube
to be the unit matrix.  For example, the Bianchi constraint for the
unit sub-cube containing $h_{11}$ is
\beq
  \mbox{Bianchi} = h_{11}g_{23}h_{62}g^\dg_{11}h_{53}g_{32} - I = 0
\eeq
Now expand the $\d$-functions in group characters
\beq
     \d[\mbox{Bianchi}] = \sum_{j=0,\oh,1,..} 
                   (2j+1) \chi_j[hghghg]
\eeq
and integrate out the $g$-variables in eq.\ \rf{Z1}.  
Displaying only terms of low order in both $h$ and $\b$, the result is
\bea
   Z &\approx& \int DV Dh \prod_{2-cubes~K}~ 
\left\{ 1 + 2\left({\b\over 4}\right)^3 \sum_{c\in K} 
\chi_\oh[(hhh)_c] \right.
\non \\
      & &\left.  + 2\left({\b\over 4}\right)^4 
                 \sum_{\stackrel{adjacent}{c_1c_2\in K}} 
                  \chi_\oh[(hhh)_{c_1}(hhh)_{c_2}] + ... \right\}
\non \\
  &\times& \exp\left[ {\b\over 2} \sum \mbox{Tr}[h] 
    + 2\left(\b\over 4\right)^{4(L-2)}\sum_{l'}
  f_{l'}^{ijkl} \mbox{Tr}[h^\dg_{ij}V_{l'}
            h^\dg_{kl}V^\dg_{l'}] \right.
\non \\
  & & \left. + 2\left({\b\over 4}\right)^{L^2} \sum_{P'}
              \mbox{Tr}[V V V^\dg  V^\dg ] \right]
\non \\
   &\approx& \int DV Dh ~ \exp\Bigl[S_L[V,h]\Bigr]
\label{Z2}
\eea
    
  At this stage, the action $S_L[V,h]$ resembles an
adjoint-Higgs Lagrangian, albeit of an unconventional form: There is an 
SU(2) gauge field $V_\m$ coupled to 24 unitary matrix-valued ``matter'' 
fields $h_{ij}$ transforming in the adjoint representation.  These matter
fields, in turn, can be subdivided into gauge-singlet fields $h_{ij,0}$,
and unit-modulus triplet fields $\vec{e}_{ij}$, where 
\beq
        h_{ij} = h_{ij,0} I + i \sqrt{1-h^2_{ij,0}} 
                    \vec{e}_{ij}\cdot \vec{\s}
\eeq
and $\vec{e}\cdot \vec{e}=1$. The unimodular $\vec{e}_{ij}$ degrees of
freedom play a role analogous to Higgs fields.  We know from the 
Elitzur theorem that the expectation values of these fields must vanish
in the absence of gauge fixing, and cannot be viewed as order parameters.
Since the coupling of the $\vec{e}_{ij}$ matter fields to 
the $V_\mu$ gauge field is very weak at large $L$, as compared to the 
self-couplings of the $\vec{e}_{ij}$ fields to each other on the 2-cubes, 
their expectation values depend primarily on these self-couplings, and on 
the complete removal of gauge redundancy in the $e$-fields through the 
choice of a unitary gauge.

   We fix to a maximal unitary gauge by first
transforming one of the 24 unimodular ``Higgs'' fields $\vec{e}_{ij}$ on
each 2-cube, denoted $\vec{e}_A$, to point in the (color) 3-direction,
i.e.
\beq
        e_{A1} = e_{A2} = 0 ~~,~~ e_{A3} = 1
\label{gf}
\eeq
This leaves a remnant U(1) symmetry, but \rf{gf} is not yet a
maximal unitary gauge fixing.  We then pick one other Higgs variable on
each 2-cube, denoted $\vec{e}_B$, and use the remaining gauge freedom
to fix
\beq
        e_{B2} = 0 ~~,~~ e_{B1} = \sin(\th_B) \ge 0
\label{gf1}
\eeq
leaving a remnant $Z_2$ symmetry.  

  In the unitary gauge [\ref{gf}-\ref{gf1}], the functional integral becomes
\beq
   Z_{ug} = \int \prod_n dV(n) \prod_{ij} dh_{ij}(n)
         \D(h_A,h_B) \d(e_{A1}) \d(e_{A2})
         \d(e_{B2}) \exp\Bigl[S_L[V,h]\Bigr]   
\eeq
where 
\bea
   \D^{-1}(h_A,h_B) &=& \int d^3\a ~ \d(e^\a_{A1}) 
                   \d(e^\a_{A2}) \d(e^\a_{B2})
\non \\
    &=& \int d^3\a ~ \d(\e_{1jk}\a_j \d_{3k}) \d(\e_{2jk}\a_j \d_{3k})
              \d(\e_{2jk}\a_j (e_{B1}\d_{1k} + e_{B3}\d_{3k}))
\non \\
   &=& e^{-1}_{B1}
\eea
From the measure
\beq
    \int dh = \int dh_0 d^3e ~\sqrt{1-h_0^2} \d(e^2-1)
\eeq
we find
\beq
\int dh_A dh_B ~ \D(h_A,h_B)  \d(e_{A1}) \d(e_{A2}) \d(e_{B2}) = 
   \int dh_{A0} dh_{B0} d\th_B \sqrt{1-h_{A0}^2}
              \sqrt{1-h_{B0}^2} \sin\th_B
\eeq
Let us take, e.g., $h_A=h_{11},~h_B=h_{44}$.  Then
\bea
Z_{ug} &=& \int \prod_n dV(n) \int dh_{11,0} dh_{44,0} d\th_{44} 
          \sqrt{1-h_{11,0}^2} \sqrt{1-h_{44,0}^2} \sin\th_{44}
\non \\
    & & \prod_{ij\ne (11),(44)} 
              dh_{ij}(n)  \exp\Bigl[S_{L}[V,h]\Bigr]   
\eea

   The final step is to integrate over the remaining $h$ degrees of
freedom in this maximal unitary gauge.  Defining $h$-expectation 
values
\bea
  \lefteqn{\langle F[h] \rangle_h ~ = ~ }
\non \\
   & &  {1\over {\Z}_h} \int dh_{11,0} dh_{44,0} d\th_{44} 
          \sqrt{1-h_{11,0}^2} \sqrt{1-h_{44,0}^2} \sin\th_{44} 
          \prod_{ij\ne (11),(44)}  dh_{ij}(n) 
\non \\
   & & \prod_{2-cubes~K}~ 
\left\{ 1 + 2\left({\b\over 4}\right)^3 \sum_{c\in K} \chi_\oh[(hhh)_c] 
       + 2\left({\b\over 4}\right)^4 
                 \sum_{\stackrel{adjacent}{c_1c_2 \in K}} 
                  \chi_\oh[(hhh)_{c_1}(hhh)_{c_2}] + ... \right\}
\non \\
     & & \times \exp\left[ {\b\over 2} \sum \mbox{Tr}[h] \right] F[h]
\eea
and
\bea
   S_{eff}[V,h] &=& 2\left(\b\over 4\right)^{4(L-2)}\sum_{l'}
  f_{l'}^{ijkl} \mbox{Tr}[h^\dg_{ij}V_{l'}h^\dg_{kl}V^\dg_{l'}]
\non \\ 
      & & + 2\left({\b\over 4}\right)^{L^2} \sum_{P'}
              \mbox{Tr}[V V V^\dg  V^\dg ] 
\eea
we have
\bea
   Z_{ug} &=& {\Z}_h \int DV \left\langle \exp\Bigl[S_{eff}[V,h]\Bigr]
                             \right\rangle_h
\non \\
          &=& {\Z}_h \int DV  \exp\Bigl[S_{eff}[V]\Bigr]
\label{zug}
\eea
where
\beq
     S_{eff}[V] =  S_{eff}[V,\langle h \rangle_h]
 + \mbox{higher-order contributions} 
\eeq
The higher-order contributions consist of next-nearest neighbor
(and more distant) couplings between $\langle h \rangle_h$ terms
on different 2-cubes, as well as closed loops in $j=1$ and higher
representations.  The large closed loops again introduce certain 
non-local correlations among the $SU(2)/Z_2$ elements of loop 
holonomies $\{U(C_i)\}$.  But
$S_{eff}[V]$ also contains local one-link terms, which provide by far 
the largest contribution to the $Z_2$-invariant part of the action.  

   For the purpose of determining saddlepoint configurations of
$S_{eff}[V]$ we may neglect the higher-order, non-local terms in
the action, so that
\bea
      S_{eff}[V] &\approx& S_{link}[V,\langle h \rangle_h] + S_{plaq}[V]
\non \\
        &=& 2\left({\b\over 4}\right)^{4(L-2)}
           \sum_{l'} f_{l'}^{ijkl}\mbox{Tr} \Bigl[
      \langle h^\dg_{ij} \rangle_h V_{l'} 
      \langle h^\dg_{kl} \rangle_h V^\dg_{l'} \Bigr]
\non \\
       & & + 2 \left({\b\over 4}\right)^{L^2}
             \sum_{P'} \mbox{Tr}[VVV^\dg V^\dg]
\label{Seff}
\eea
and, for this particular gauge choice, we find
\beq
     \begin{array}{lcl}
   \langle h_{11} \rangle_h = {\b\over 4} I 
           + i {8\over 3\pi} \s_3   
&~,~&
   \langle h_{12} \rangle_h = {\b\over 4} I 
           - i \left({\b\over 4}\right)^8 {8\over 3\pi} \s_3   
\cr
   \langle h_{13} \rangle_h = {\b\over 4} I 
           - i \left({\b\over 4}\right)^8 {8\over 3\pi} \s_3   
&~,~&
   \langle h_{14} \rangle_h = {\b\over 4} I 
           - i \left({\b\over 4}\right)^8 {2\over 3} \s_1
\cr
   \langle h_{41} \rangle_h = {\b\over 4} I 
           - i \left({\b\over 4}\right)^8 {8\over 3\pi} \s_3   
&~,~&
   \langle h_{42} \rangle_h = {\b\over 4} I 
           - i \left({\b\over 4}\right)^8 {2\over 3} \s_1   
\cr
   \langle h_{43} \rangle_h = {\b\over 4} I 
           - i \left({\b\over 4}\right)^8 {2\over 3} \s_1 
&~,~& 
   \langle h_{44} \rangle_h = {\b\over 4} I 
           + i {2\over 3} \s_1  
\cr
   \langle h_{21} \rangle_h = {\b\over 4} I 
           - i \left({\b\over 4}\right)^8 {2\over 3} \s_1
&~,~&
   \langle h_{22} \rangle_h = {\b\over 4} I 
           - i \left({\b\over 4}\right)^4 {2\over 3} \s_1  
\cr
   \langle h_{23} \rangle_h = {\b\over 4} I 
           - i \left({\b\over 4}\right)^8 {8\over 3\pi} \s_3   
&~,~&
   \langle h_{24} \rangle_h = {\b\over 4} I 
           - i \left({\b\over 4}\right)^8 {2\over 3} \s_1
\cr 
   \langle h_{51} \rangle_h = {\b\over 4} I 
           - i \left({\b\over 4}\right)^8 {8\over 3\pi} \s_3
&~,~&
   \langle h_{52} \rangle_h = {\b\over 4} I 
           - i \left({\b\over 4}\right)^8 {2\over 3} \s_1  
\cr
   \langle h_{53} \rangle_h = {\b\over 4} I 
           - i \left({\b\over 4}\right)^4 {8\over 3\pi} \s_3 
&~,~&
   \langle h_{54} \rangle_h = {\b\over 4} I 
           - i \left({\b\over 4}\right)^8 {8\over 3\pi} \s_3 
\cr
   \langle h_{31} \rangle_h = {\b\over 4} I 
           - i \left({\b\over 4}\right)^8 {2\over 3} \s_1  
&~,~&
   \langle h_{32} \rangle_h = {\b\over 4} I 
           - i \left({\b\over 4}\right)^4  {2\over 3} \s_1 
\cr
   \langle h_{33} \rangle_h = {\b\over 4} I 
           - i \left({\b\over 4}\right)^8 {2\over 3} \s_1
&~,~&
   \langle h_{34} \rangle_h = {\b\over 4} I 
           - i \left({\b\over 4}\right)^8 {8\over 3\pi} \s_3   
\cr
   \langle h_{61} \rangle_h = {\b\over 4} I 
           - i \left({\b\over 4}\right)^8 {8\over 3\pi} \s_3 
&~,~&
   \langle h_{62} \rangle_h = {\b\over 4} I 
           - i \left({\b\over 4}\right)^8 {2\over 3} \s_1 
\cr
   \langle h_{63} \rangle_h = {\b\over 4} I 
           - i \left({\b\over 4}\right)^4 {8\over 3\pi} \s_3
&~,~&
   \langle h_{64} \rangle_h = {\b\over 4} I 
           - i \left({\b\over 4}\right)^8 {8\over 3\pi} \s_3   
\end{array} 
\label{h}
\eeq
Inserting \rf{h} into $S_{link}$ we have, to leading order in $\b$,
\bea
  \lefteqn{ S_{link}[V,\langle h \rangle_h] = 
     2\left({\b\over 4}\right)^{4(L-2)} \times \sum_n}
\non \\
   & & \left\{
       \left({8\over 3\pi}\right)^2 \left({\b \over 4}\right)^8
        \mbox{Tr}[\s_3 V_z(n) \s_3 V^\dg_z(n)]
     + \left({2\over 3}\right)^2 \left({\b \over 4}\right)^8
        \mbox{Tr}[\s_1 V_z(n) \s_1 V^\dg_z(n)] \right.
\non \\
   & & - \left({8\over 3\pi}\right)^2 \left({\b \over 4}\right)^{12}
        \mbox{Tr}[\s_3 V_y(n) \s_3 V^\dg_y(n)]
     - \left({2\over 3}\right)^2 \left({\b \over 4}\right)^{12}
        \mbox{Tr}[\s_1 V_y(n) \s_1 V^\dg_y(n)]
\non \\
   & &  - \left({8\over 3\pi}\right)^2 \left({\b \over 4}\right)^{12}
        \mbox{Tr}[\s_3 V_x(n) \s_3 V^\dg_x(n)]
     - \left({2\over 3}\right)^2 \left({\b \over 4}\right)^{12}
        \mbox{Tr}[\s_1 V_x(n) \s_1 V^\dg_x(n)]  
\non \\
   & &   + ~ \mbox{const.} \Bigr\}
\label{Slink}
\eea
Each term in $S_{link}$ is proportional to a component of
a $V$-link variable in the adjoint representation, and is insensitive
to the center degrees of freedom.
The spatial asymmetry of $S_{link}$ in \rf{Slink} is, of course, 
due to the particular unitary gauge choice.\footnote{Despite this asymmetry, 
the expectation value of any Wilson loop on the V-lattice, 
evaluated in the full (gauge-dependent) effective action $S_{eff}[V]$ 
defined in eq.\ \rf{zug}, is necessarily independent of the gauge choice.
This should be clear from the construction, where a gauge-invariant action
$S_L[V,h]$ is gauge-fixed, followed by integration over the remaining
$h$ degrees of freedom.}

   We now look for saddlepoints of $S_{eff}$.  It can be seen from inspection 
of \rf{Slink} that $S_{link}$ is maximized by
\beq
   V_x(n) = i\s_2 Z_x(n)  ~~,~~  V_y(n) = i\s_2 Z_y(n)  ~~,~ 
   V_z(n) = Z_z(n)  
\label{ground}
\eeq
where the $Z_\m(n) = \pm I$ are center elements.  The $S_{plaq}$
term is also maximized if the $Z_\m$ link variables are gauge-equivalent
to the identity under the remnant $Z_2$ gauge symmetry. 
With this choice the effective action $S_{eff} \approx S_{link}+S_{plaq}$
is maximized, and the configuration \rf{ground} is the ground state, 
gauge equivalent to the identity.  The fact that this ground state
is unique, up to a $Z_2$ gauge transformation, is again due to the
maximal unitary gauge fixing.

   Now consider the configuration
\bea
      V_y(\vec{n}) &=& \left\{ \begin{array}{rl}
                    -i\s_2 & n_1 \ge 2, ~ n_2=1 \cr 
                    +i\s_2 & \mbox{otherwise} \end{array} \right.
\non \\
     V_x(\vec{n}) &=& i\s_2 
\non \\
     V_z(\vec{n}) &=& I 
\label{vortex}
\eea
This configuration is a center vortex, one lattice spacing
thick, running in the $z$-direction.  It is not hard to see that
this configuration, like the trivial ground state, is also
a saddlepoint of $S_{eff}$.  In the first place, \rf{vortex}
is a global maximum of $S_{link}$, since the thin vortex configuration
\rf{vortex} differs from link variables in the ground state only 
by center elements, to which $S_{link}$ is insensitive.  Secondly,
this configuration is also a stationary point of the 
plaquette action $S_{plaq}$ \cite{Yoneya}.  This is because a plaquette at one
of its extremal values $\oh\mbox{Tr}[VVV^\dg V^\dg] = \pm 1$ 
varies at most quadratically,
and is therefore stationary, with respect to a small variation $\d V$
of any link respecting the unitarity constraint $(V+\d V)(V+\d V)^\dg = I$.
All of the plaquettes formed from \rf{vortex} are extremal. 
So the center vortex \rf{vortex} is certainly a stationary configuration
of $S_{eff}$, the remaining question is whether it is stable;
i.e.\ whether the vortex is a local \emph{maximum} of $S_{eff}$, in
which case it is a stable saddlepoint.

   The stability issue is settled by looking at the eigenvalues of
\beq
     {\d^2 S_{eff} \over \d V_\m(n_1) \d V_\n(n_2)} =
        {\d^2 S_{link} \over \d V_\m(n_1) \d V_\n(n_2)}
      + {\d^2 S_{plaq}  \over \d V_\m(n_1) \d V_\n(n_2)} 
\eeq
where, from the coefficients shown in eqs.\ \rf{Seff} and \rf{Slink},
we see that
\beq
      {\d^2 S_{link}  \over  \d V_\m(n_1) \d V_\n(n_2)}
         \sim \left({\b\over 4}\right)^{4(L-2)+12}
      ~~~,~~~
      {\d^2 S_{plaq}  \over  \d V_\m(n_1) \d V_\n(n_2)}
         \sim \left({\b\over 4}\right)^{L^2}
\eeq
The crucial observation is that for $\b/4 \ll 1$ and
\beq
      4(L-2) + 12 < L^2
\label{L}
\eeq
the contribution of $S_{plaq}$ to the stability matrix (and therefore
to the eigenvalues of the stability matrix) is negligible compared to
the contribution of $S_{link}$, which has only stable modes.  Thus from 
the fact that the thin center vortex
\rf{vortex} is both a stationary point of $S_{eff}$, and a stable
saddlepoint of $S_{link}$, we can conclude that the vortex is also a
stable saddlepoint of the full effective action $S_{eff}$ when
condition \rf{L} is satisfied.

   Condition \rf{L} is satisfied for $L \ge 5$ lattice spacings.  It
is probably no coincidence that this is also where the adjoint string
breaks in strong-coupling lattice gauge theory (as can be easily
verified from looking at correlations of Polyakov lines in the adjoint
representation).  It has been known for a long time that, at
intermediate distance scales and weak couplings, the static
quark-antiquark potential is roughly (and maybe even accurately
\cite{Bali}) proportional to the quadratic Casimir of quark color
representation.  In ref.\ \cite{name} this phenomenon was dubbed
``Casimir scaling,'' and the problems it poses for monopole and vortex
theories was discussed.  In ref.\ \cite{Cas} we have argued that the
problems with respect to the vortex theory can in principle be
resolved by taking into account the finite thickness of the vortex,
which should be comparable to the adjoint string-breaking length.  A
result of the analysis carried out above is that there are stable
center vortices, one lattice spacing thick on the V-lattice,
corresponding to $L \ge 5$ on the original U-lattice.  This gives an
estimate for the vortex thickness of $L=4$ lattice spacings in
U-lattice units.  This vortex thickness happens to be exactly the
length where adjoint string-breaking occurs in the strong-coupling
Wilson action, and is therefore consistent with the reasoning in ref.\
\cite{Cas}.

   By inspection of $S_{eff}$, we see that the action of a center
vortex configuration in D=3 dimensions (ignoring any further quantum
corrections) is
\beq
      8\left({\b\over 4}\right)^{L^2} ~\times~ \mbox{vortex length} 
\label{av}
\eeq
on the V-lattice.
On the other hand, the entropy of a linelike
object (essentially the entropy of a random walk) is a constant of
$O(1)$ times the line length.  Thus, center vortices are stable
saddlepoints of the long range effective action whose entropy per unit
length greatly exceeds their action per unit length.  These objects
therefore percolate throughout the lattice volume, and confine color
charge in any half-integer group representation.

    This picture has been derived in a particular unitary gauge.   
A natural question is whether the saddlepoints of the effective action
would be qualitatively different had 
we chosen to gauge-fix, instead of $\vec{e}_{11}$ and $\vec{e}_{44}$,
some other plaquette variables (or combination of plaquette variables)
on the 2-cubes.  Although an analysis of all possible unitary gauge choices
is beyond us at present, it is easy to see that center vortices
must be stable saddlepoints in a very large class of gauges.  
We first note that, by definition, any maximal unitary gauge must completely 
determine the minimal action configuration of the $V_\m$ fields, up 
to residual $Z_2$ gauge transformations.  Then a sufficient condition 
for center vortex stability is simply that
the classical ground state of $S_{eff}[V]$ has the 
form of a pure gauge $V_\m(x) = g(x)g^{-1}(x+\hat{\m})$, and that this
is also a maximum of the center-invariant $S_{link}$ 
part of the effective action.  
In that case, the effective action must have stable thin vortex solutions
at large $L$.  This is because the stable fluctuation modes around a 
thin vortex, associated with the center-insensitive $S_{link}$ term, will 
overwhelm (at sufficiently large $L$) any unstable modes associated 
with $S_{plaq}$.  The vortex action at a given $L$ will always have the 
value shown in eq.\ \rf{av} above, so the entropy of the 
configuration will exceed the action at strong couplings. 

   Our findings for the strong-coupling theory do not, however, prove that
vortices also dominate the vacuum at weaker couplings; the strong and
weak coupling regimes are separated by a roughening phase transition,
and this transition prevents a simple extrapolation from one regime to
the other.  The result is, nonetheless, significant for continuum
physics in two ways: First, it supports the very general argument that
if the asymptotic quark potential is sensitive only to $N$-ality, then
the confining field configurations must be center vortices.  Secondly,
it provides an explicit illustration of how center vortices are
stabilized, via color-screening (center-invariant) 
terms in the long-range effective action.  
  
  Although strong-coupling methods only apply at strong couplings, the
general approach we have advocated here should extend, at least in
principle, to weak-coupling lattice gauge theory.  The central idea
is that if we want to extract a \emph{local} long-range effective action
from the Wilson action, then it is necessary to include composite operators, 
transforming like adjoint matter fields, in the derivation.
With this motivation we define gluelump operators
\beq
     G_M[x;U] = \sum_{C_x} a_M(C_x) \prod_{l\in C_x} U_l
\eeq
which, coupled to a static adjoint source at site $x$, create gauge-invariant
eigenstates of the appropriate transfer matrix.
The $C_x$ are paths on the lattice beginning and ending at 
$x$.\footnote{At strong couplings, the only relevant paths $C_x$ run 
around single plaquettes.} The index $M$ specifies the time (i.e. 
worldline) direction of the static source, and any other (e.g.\ spin) 
degeneracies.
The transformation from a pure gauge theory on a fine lattice,
to a theory of gauge fields $V_\m$ coupled to gluelump 
fields $H_M$ on a coarse lattice, could then be accomplished as follows:
\beq
     \exp\Bigl[ S_{eff}[V,H] \Bigr] = 
          \int DU \prod_{l'}\d\Bigl(V_{l'}-Q_{l'}[U]\Bigr)
            \prod_{M,x'}\d\Bigl(H_M(x')-G_M[x';U]\Bigr) e^{S_W[U]}
\eeq
where $x',l'$ denote sites and links on the coarse lattice. The expression
$Q_{l'}[U]$ represents a suitable ``fat link'' function, i.e.\ a 
superposition of Wilson lines on the fine lattice which run between 
sites bounding link $l'$ on the coarse lattice.  The constraints imposed
by the delta-functions can be softened by replacing delta-functions with 
exponentials, as in the ``perfect action'' approach \cite{Has}.  
The end result of this procedure
will be an effective long-range action consisting of gauge fields coupled 
to a set of adjoint Higgs-like fields.  Possibly this scheme
can be implemented numerically at moderately weak couplings, along the lines 
of the Monte Carlo renormalization-group.

    In the strong-coupling analysis carried out in this article, 
we have seen how color-screening, center-invariant ``Higgs'' terms 
predominate in the long-range effective action beyond the
adjoint string-breaking scale, and stabilize center vortex configurations. 
The entropy of these configurations exceeds the cost in action, and vortex
configurations percolate throughout the lattice.  
We think it likely that these important 
features are not specific to strong couplings,
and also characterize the effective action of lattice QCD in the
continuum limit.

\acknowledgments{ Our research is supported in part by
Fonds zur F\"orderung der
Wissenschaftlichen Forschung P13997-PHY (M.F.), the U.S. Department of 
Energy under Grant No.\ DE-FG03-92ER40711 (J.G.), 
the ``Action Austria-Slovakia: Cooperation in Science and Education''
(Project No.\ 30s12) and the Slovak Grant 
Agency for Science, Grant No. 2/7119/2000 (\v{S}.O.). }

\end{document}